\documentclass[12pt,a4paper,reqno]{article}

\usepackage{amsmath,latexsym,amssymb,amsthm}
\usepackage{graphicx}

\usepackage{amsmath}\usepackage{amssymb}
\usepackage{amsthm}
\newcommand{\head}{\text{head}}
\newcommand{\tail}{\text{tail}}
\newcommand{\ok}{\text{ok}}
\newcommand{\fail}{\text{fail}}

\newcommand{\semilinespace}{\vspace{0.5\baselineskip}}

\newcommand{\be}{\begin{equation}}
\newcommand{\ee}{\end{equation}}
\renewcommand{\(}{\left(}
\renewcommand{\)}{\right)}

\newcommand{\1}{\mathbf{1}}

\renewcommand{\S}{\mathcal{S}}

\renewcommand{\H}{\mathcal{H}}

\newcommand{\ox}{\otimes}

\newcommand{\<}{\langle}
\renewcommand{\>}{\rangle}
\newcommand{\half}{\tfrac{1}{2}}

\newcommand{\third}{\tfrac{1}{3}}

\newcommand{\FR}{Frauchiger and Renner}

\theoremstyle{definition}

\theoremstyle{remark}

\numberwithin{equation}{section}

\newcommand{\secref}[1]{Section~\ref{#1}}

\begin{document}

\title{\bf{Single-world theory of the extended Wigner's friend experiment}}

\author{Anthony Sudbery$^1$\\[10pt] \small Department of Mathematics,
University of York, \\[-2pt] \small Heslington, York, England YO10 5DD\\
\small $^1$ as2@york.ac.uk}

\date{14 November 2016}

\maketitle

\begin{abstract}

 Frauchiger and Renner have recently claimed to prove that ``Single-world interpretations of quantum theory cannot be self-consistent". This is contradicted by a construction due to Bell, inspired by Bohmian mechanics, which shows that any quantum system can be modelled in such a way that there is only one ``world" at any time, but the predictions of quantum theory are reproduced. This Bell-Bohmian theory is applied to the experiment proposed by Frauchiger and Renner, and their argument is critically examined. It is concluded that it is their version of ``standard quantum theory", incorporating state vector collapse upon measurement, that is not self-consistent. 

\end{abstract}

\section{Introduction}

In 1984 John Bell \cite{Bell:beables} proposed an interpretation of quantum field theory in which certain field variables always have definite values. This can be generalised to any quantum system \cite{QMPN}, giving a theory in which any chosen set of commuting observables --- the \emph{beables} of the theory --- always have definite values, and yet the results of measurements are always distributed as predicted by quantum mechanics. Bell's theory was an extension of Bohmian quantum mechanics.

Recently Frauchiger and Renner have declared that this is impossible \cite{FrauchigerRenner}. They describe an experiment, the ``extended Wigner's friend experiment", in which, they claim, the predictions of quantum mechanics and the assumption that each measurement in the experiment has a unique result, together lead to a contradiction. In this paper we examine the Bell-Bohmian description of the extended Wigner's friend experiment in an attempt to identify the source of this contradiction.

The contents of the paper are as follows. Section 2 is an outline of the Bell-Bohmian theory. Section 3 contains a description of the extended Wigner's friend experiment and a summary of the argument of Frauchiger and Renner. In Section 4 we analyse the experiment in terms of Bell-Bohmian theory and show how it avoids the contradiction found by Frauchiger and Renner. Section 5 contains discussion, leading to the conclusion that the source of the contradiction is the use of the projection postulate for measurements by different agents.

\section{Bell-Bohmian theory}\label{BellBohm}

This interpretation was inspired by Bohm's interpretation of non-relativistic many-particle quantum mechanics (see e.g. \cite{Bohmian} p. 145), according to which particles always have definite positions. The motion of the particles is governed deterministically by the wave function, which thus has the role of a force acting on the system rather than a description of the state of the system. To emphasise this role, Bell \cite{Bell:pilot} calls the wave function a ``pilot wave". This evolves according to the time-dependent Schr\"odinger equation.

In Bell's generalisation of this interpretation (\cite{Bell:beables}, \cite{QMPN} p. 215), the many-particle system can be replaced by any quantum system S, with states described by vectors in a Hilbert space $\mathcal{S}$, and the positions of the particles replaced by any set of commuting variables, which are known as \emph{beables}. These are taken to have definite values, so the actual real state of the system is described by a state vector in one of the simultaneous eigenspaces $\S_i$ of the beables (which are also known \cite{QMPN} as \emph{viable} subspaces). The evolution of this state is governed by another time-dependent vector, the \emph{pilot vector} $|\Psi\>\in \S$, which satisfies the time-dependent Schr\"odinger equation with the Hamiltonian $H$ determined by the physics of the system. This pilot vector can be decomposed into its components in the viable subspaces $\S_i$:
\[
|\Psi(t)\> = \sum_i |\psi_i(t)\> \qquad \text{with} \quad |\psi_i(t)\> \in \S_i,
\]
and the real state at time $t$ is taken to be one of the components $|\psi_i(t)\>$.

The real state changes in time, not deterministically as in the original Bohmian mechanics, but stochastically: it makes transitions between the preferred subspaces $S_i$ with transition probabilities given by 

\semilinespace

{\bf Bell's postulate}: The real state of the system is one of the components of the pilot state vector in one of the viable subspaces $\S_i$. If, at time $t$, the real state is the component $|\psi_i(t)\>\in\S_i$, then the probability that at time $t + \delta t$ the real state is $|\psi_j(t + \delta t)\>\in\S_j$ is $w_{ij}\delta t$ where the transition probability $w_{ij}$ is given by
\be\label{Bell}
w_{ij} = \begin{cases} \frac{2\text{Re}[(i\hbar)^{-1}\<\psi_j(t)|H|\psi_i(t)\>]}{\<\psi_i(t)|\psi_i(t)\>} &\text{if this is } \ge 0\\
                       0                                                  &\text{if it is negative}
         \end{cases}
\ee
It follows from this \cite{QMPN}\footnote{The proof in \cite{QMPN} refers to a slightly  different, and less satisfactory, form of Bell's postulate, but it is easily adapted so as to apply to the form given here.} that the probability $p_i(t) $ that the real state of the system at time $t$ is $|\psi_i(t)\>$ is given by the Born rule ($p_i(t) = \<\psi_i(t)|\psi_i(t)\>$) at all positive times $t$, if the probabilities are so given at the initial time $t = 0$.

This framework can be generalised still further \cite{BacciaDickson, verdammte} to allow for the possibility that the viable subspaces $\S_i$ vary with time; it then includes the modal interpretation of quantum mechanics.

Although this theory is indeterministic, it can be shown \cite{determlimit, Vink} that Bohm's deterministic theory can be obtained as a continuum limit of Bell's original theory of the above form, in which he took the points of space to be a discrete lattice.

\section{The extended Wigner's friend experiment}

This section contains a description of the experiment designed by Frauchiger and Renner \cite{FrauchigerRenner} to demonstrate that any theory which is compliant with quantum theory and describes a single world cannot be self-consistent. After describing the experiment, we will outline the argument of Frauchiger and Renner for this conclusion.

The experiment contains two experimenters $F_1$ and $F_2$ (Wigner's friends), who perform experiments on two two-state quantum systems, a coin $C$ with orthonormal basis states $|\head\>_C$ and $|\tail\>_C$, and an electron $S$ with spin states $|\uparrow\>_S$ and $|\downarrow\>_S$; it also contains Wigner $W$ and his assistant $A$, who can perform measurements on $F_1$ and $F_2$ as well as the coin $C$ and the electron spin $S$. Irrelevant degrees of freedom of the four experimenters are suppressed, so each of them is regarded as having just two independent states, which record the results of their measurements.


Before the experiment starts the coin is prepared in the state 
$\sqrt{\third}|\head\> + \sqrt{\tfrac{2}{3}}|\tail\>$. 

At time $t = 0$ experimenter $F_1$ observes the coin and records the result $r =$ ``head" or ``tail", thereby being put into a memory state $|r\>_{F_1}$.

At time $t = 1$, $F_1$ prepares the electron as follows: if the result of the measurement at $t = 0$ was $r =$ ``head", $F_1$ prepares the electron in spin state $|\downarrow\>_S$; if $r = $ ``tail", they prepare it in spin state $|\rightarrow\>_S = \tfrac{1}{\sqrt{2}}\big(|\uparrow\>_S + |\downarrow\>_S$. 

At time $t = 2$ experimenter $F_2$ measures the spin $z\half\hbar$ of the electron $(z = \pm)$ in the basis $\{|\uparrow\>, |\downarrow\>\}$ and records the result, thereby being put into a memory state $|z\>_{F_2}$.

At time $t = 3$ Wigner's assistant $A$ measures $F_1$, together with the coin, in the basis
\begin{align*}
|\ok\>_{F_1C} &= \tfrac{1}{\sqrt{2}}\big(|\head\>_{F_1}|\head\>_C - |\tail\>_{F_1}|\tail\>_C\big)\\
|\fail\>_{F_1C} &= \tfrac{1}{\sqrt{2}}\big(|\head\>_{F_1}|\head\>_C + |\tail\>_{F_1}|\tail\>_C\big),
\end{align*}
and records the result $x =$ ``ok" or ``fail".

At time $t = 4$ Wigner measures $F_2$, together with the electron, in the basis
\begin{align*}
|\ok\>_{F_2S} &= \tfrac{1}{\sqrt{2}}\big(|-\>_{F_2}|\downarrow\>_S - |+\>_{F_2}|\uparrow\>_S\big)\\
|\fail\>_{F_2S} &= \tfrac{1}{\sqrt{2}}\big(|-\>_{F_2}|\downarrow\>_S + |+\>_{F_1}|\uparrow\>_S\big),
\end{align*}
and records the result $w =$ ``ok" or ``fail".

At the end of the experiment Wigner and his assistant compare the results of their measurements. They repeat the experiment again and again, stopping when they find $x = w = ``\ok"$. The question is whether it is possible for the procedure to stop.

Frauchiger and Renner argue as follows. Let us assume that the experiment is described by a theory $T$ with the following three properties:

{\bf QT} \emph{Compliance with quantum theory}: $T$ forbids all experimental results that are forbidden by standard quantum theory.

{\bf SW} \emph{Single world}: $T$ rules out the occurrence of more than one single outcome if an experimenter measures a system once.

{\bf SC} \emph{Self-consistency}: $T$'s statements about measurement outcomes are logically consistent (even if they are obtained by considering the perspectives of different experimenters).\footnote{This is the formulation of Frauchiger and Renner. Elsewhere they state that this property ``demands that the laws of a theory $T$ do not contradict each other". These are not the same. If the laws of a theory $T$ contradicted each other, then $T$ simply would not exist as a theory. But as stated here, {\bf SC} is not a very interesting requirement: there is no logical reason why statements existing in different perspectives should be consistent (think of statements about the order of events in different frames of reference, in special relativity). However, we show in this paper that even in this form there is no contradiction between {\bf QT}, {\bf SW} and {\bf SC}.}

Then we have the following implications:

{\bf 1.} Suppose that $F_1$, in the measurement at $t = 1$, gets the result $r =$ ``tail". Then $F_1$ prepares the electron spin $S$ in the state $|\rightarrow\>_S$. When $F_2$ measures $S$ at $t = 2$, $F_2$ and $S$ are put into the entangled state $|\fail\>_{F_2S}$. This is not affected by $A$'s measurement of $F_1C$ at $t = 3$, so $W$, on measuring $F_1C$ at $t = 4$, will, by {\bf QT}, get the result $w =$ ``fail". Thus
\be\label{r implies w}
r(1) = \tail\; \Longrightarrow \; w(4) = \fail 
\ee
(this allows for the possibility that the values of $r$ and $w$ might vary with time). Since, by {\bf SW}, the value of $r(1)$ must be either ``head" or ``tail",  and the value of $w(4)$ must be either ``ok" or ``fail", it follows that 
\be\label{w implies r}
w(4) = \ok \; \Longrightarrow \; r(1) = \head.
\ee

{\bf 2.} Suppose, on the other hand, that $F_1$ gets the result $r =$ ``head" at $t = 1$. Then the state of the electron spin after this measurement must be $|\downarrow\>$. Hence, by {\bf QT}, $F_2$, in the measurement at $t = 2$, must get the result $z = -$. Thus
\be\label{r implies z}
r(1) = \head \; \Longrightarrow \; z(2) = -.
\ee

{\bf 3.} Now consider $F_2$'s measurement of $z$ at $t = 2$. After $F_1$'s preparation of the electron spin, the state of $F_1$, the coin and the electron is
\[
\sqrt{\third}|\head\>_{F_1C}|\downarrow\>_S + \sqrt{\tfrac{2}{3}}|\tail\>_{F_1C}|\rightarrow\>_S = \sqrt{\third}|\tail\>_{F_1C}|\uparrow\>_S + \sqrt{\tfrac{2}{3}}|\fail\>_{F_1C}|\downarrow\>_S.
\]
Hence if the result of $F_2$'s measurement of $S$ is $z = -$, then the result of $A$'s measurement of $F_1C$ at $t = 3$ must be $x =$ ``fail":
\be\label{z implies x}
z(2) = - \; \Longrightarrow \; x(3) = \fail.
\ee

{\bf 4.} After $F_2$'s measurement of the electron spin, the state of $F_1$ and $F_2$ (and their laboratories) is
\begin{multline*}
\sqrt{\third}|\tail\>_{F_1}|\tail\>_C|+\>_{F_2}|\uparrow\>_S + \sqrt{\tfrac{2}{3}}|\fail\>_{F_1C}|-\>_{F_2}|\downarrow\>_S\\
= \tfrac{1}{2\sqrt{3}}\big(|\ok\>_{F_1C}|\ok\>_{F_2S} - |\ok\>_{F_1C}|\fail\>_{F_2S} + |\fail\>|\ok\>_{F_2S}\big) + \tfrac{\sqrt{3}}{2}|\fail\>_{F_1C}|\fail\>_{F_2S}.
\end{multline*}
This has non-zero coefficient of $|\ok\>_{F_1C}|\ok\>_{F_2S}$, so
\[
x(4) = w(4) = \ok \text{ is possible.}
\]
But $W$'s measurement of $F_2S$ does not affect the state of $A$, so $x(4) = x(3)$. Thus in the measurements of $A$ and $W$ at $t = 3$ and $4$,
\be\label{x and w}
x(3) = w(4) = \ok \text{ is possible.}
\ee

Now we have
\begin{align*}
w(4) = \ok \; &\Longrightarrow \; r(1) = \head \qquad \text{by \eqref{w implies r}}\\
           &\Longrightarrow \; z(2) = - \qquad\qquad \text{by \eqref{r implies z}}\\
           &\Longrightarrow \; x(3) = \fail \quad \text{by \eqref{z implies x}}
\end{align*}
which contradicts \eqref{x and w}. Frauchiger and Renner conclude that no theory can have all three properties {\bf SW}, {\bf QT} and $\bf SC$.
\section{Bell-Bohmian theory of the experiment}\label{Bell-Bohm}

Bell-Bohmian theory assumes a pilot vector in the Hilbert space of the whole experiment, evolving purely according to the unitary operator describing the dynamics (i.e. with no application of the projection postulate after measurements). In this it resembles Everettian quantum mechanics, but the metaphysical interpretation is different, as described in \secref{BellBohm}. The Hilbert space in question is
\[
\H_{F_1}\otimes\H_{F_2}\otimes\H_A\otimes\H_W\otimes\H_C\otimes\H_S
\]
where $\H_C$ and $\H_S$ are two-dimensional, with orthonormal bases $\{|\head\>,|\tail\>\}$ and $\{|\uparrow\>, |\downarrow\>\}$ respectively; and $\H_{F_1}, \H_{F_2}, \H_A$ and $\H_W$ are all 3-dimensional, with bases labelled by $r, z, x$ and $w$, each taking the two values described in Section 2 and also a third value $0$ to describe the ``ready" state of the observer before making any measurement. We take $r, z, x$ and $w$ to be the beables of the system, which always have definite values. Thus the real state vector of the system always lies in one of the 81 viable subspaces
\[
|r\>_{F_1}|z\>_{F_2}|x\>_A|w\>_W\otimes\H_C\otimes\H_S.
\]
and is one of the projections of the pilot vector onto these subspaces.

  In order to analyse the experiment, we need to be more precise about the way in which $F_1$ prepares the spin state after the coin toss at $t = 0$. I will assume that before the coin toss, the electron spin is prepared in some known initial state $|0\>_S\in\H_S$; after the coin toss, $F_1$ applies to the electron either a unitary operator which takes $|0\>$ to $|\downarrow\>$ or one which takes $|0\>$ to $|\rightarrow\>$, according to the result of the toss. Then the real state vector before the experiment starts is the same as the pilot state, namely
\[
 |0\>_{F_1}|0\>_{F_2}|0\>_A|0\>_W\(\sqrt{\third}|\head\>_C + \sqrt{\tfrac{2}{3}}|\tail\>_C\)|0\>_S. 
\]
  	At $t=0$, after $F_1$'s measurement of the coin, the pilot vector becomes 
\[
|\Psi(0)\> = \left(\sqrt{\third}|\head\>_{F_1C} + \sqrt{\tfrac{2}{3}}|\tail\>_{F_1C}\right)|0\>_S|0\>_{F_2}|0\>_A|0\>_W,
\]
where $|\head\>_{F_1C} = |\head\>_{F_1}|\head\>_C$ and similarly for ``tail",
but the real state vector is one of the two summands in this. We will consider
\[
|\Phi(0)\> = |\tail\>_{F_1C}|0\>_S|0\>_{F_2}|0\>_A|0\>_W.
\]
At $t =1$, after $F_1$ has prepared the electron spin, the pilot state is
\[
|\Psi(1)\> = \left(\sqrt{\third}|\head\>_{F_1C}|\downarrow\>_S + \sqrt{\tfrac{2}{3}}|\tail\>_{F_1C}|\rightarrow\>_S\right)|0\>_{F_2}|0\>_A|0\>_W
\]
The real state is one of the two summands in $|\Psi(1)\>$; we take
\[
|\Phi(1) = \sqrt{\tfrac{2}{3}}|\tail\>_{F_1C}|\rightarrow\>_S|0\>_{F_2}|0\>_A|0\>_W.
\]
After $F_2$'s measurement of $S$ at $t = 2$, the pilot state becomes
\[
|\Psi(2)\> = \sqrt{\tfrac{1}{3}}\bigg(|\head\>_{F_1C}|-\>_{F_2S} + |\tail\>_{F_1C}|+\>_{F_2S} + 
|\tail\>_{F_1C}|-\>_{F_2S}\bigg)|0\>_A|0\>_W,
\]
which has three components with definite values of $r, z, x$ and $w$ (\emph{viable} components), one of which is
\[
|\Phi(2)\> = \sqrt{\tfrac{1}{3}}|\tail\>_{F_1C}|+\>_{F_2S}|0\>_A|0\>_W.
\]
After $A$'s measurement of $F_1$ and $C$ at $t = 3$, the pilot vector becomes
\begin{multline*}
|\Psi(3)\> = \bigg(\sqrt{\tfrac{1}{6}}\Big( - |\ok\>_{F_1C}|\ok\>_A + |\fail\>_{F_1C}|\fail\>_A\Big)|+\>_{F_2S}\>\\
+ \sqrt{\tfrac{2}{3}}|\fail\>_{F_1C}|\fail\>_A|-\>_{F_2S}\bigg)|0\>_W
\end{multline*}
which has six viable components, one of which is
\[
|\Phi(3)\> = \sqrt{\tfrac{1}{12}}|\tail\>_{F_1C}|+\>_{F_2S}|\ok\>_A|0\>_W.
\]
After $W$'s measurement of $F_2$ and $S$ at $t = 4$, the pilot vector becomes
\begin{multline*}
|\Psi(4)\> = \sqrt{\tfrac{1}{12}}\Big(|\ok\>_{F_1C}|\ok\>_A + |\fail\>_{F_1C}|\fail\>_A\Big)|\ok\>_{F_2S}|\ok\>_W\\
+ \sqrt{\tfrac{1}{12}}\Big(- |\ok\>_{F_1C}|\ok\>_A + 3|\fail\>_{F_1C}|\fail\>_A\Big)|\fail\>_{F_2S}|\fail\>_W
\end{multline*}
which has sixteen viable components, one of which is
\[
|\Phi(4)\> = -\sqrt{\tfrac{1}{24}}|\tail\>_{F_1C}|-\>_{F_2S}|\ok\>_A|\ok\>_W.
\]

According to Bell-Bohmian theory, at all times Wigner, his assistant and his two friends are in a single world with definite values of $r, z, x$ and $w$, the results of their measurements. But Frauchiger and Renner argue that this leads to the contradictory implications \eqref{r implies w}, \eqref{r implies z}, \eqref{z implies x} and \eqref{x and w}. We will show, on the contrary, that in Bell-Bohmian theory it is possible that the real state undergoes the transitions
\[
|\Phi(0)\> \longrightarrow |\Phi(1)\> \longrightarrow |\Phi(2)\> \longrightarrow |\Phi(3)\> \longrightarrow\> |\Phi(4)\>.
\]
It follows that in this theory the implication \eqref{r implies w} ($r(1) = \tail\; \Longrightarrow \; w(4) = \fail$) does not hold: it is possible for $F_1$ to get the result $r = $ ``tail" (and, incidentally, to remain in a state registering this result) while $W$ gets the result $w = $``ok".
 
To establish this, we will need to see what transitions between viable states are allowed by Bell's postulate, and for this we need a model of the processes by which the measurements are made. The following is a general theory of such a process. We consider an experimenter $E$ measuring an observable $X$ on a system $S$, whose basis of eigenstates of $X$ is $\{|1\>_S,|2\>_S\}$, and suppose that the process takes place as follows. The relevant states of the experimenter are taken to be $|0\>_E,|1\>_E,|2\>_E$, where $|0\>_E$ is the state of the experimenter before the measurement, and $|1\>_E$ and $|2\>_E$ are the states of the experimenter registering the results $X = 1$ and $X = 2$. In the course of the measurement the joint state $|1\>_S|0\>_E$ evolves to $|1\>_S|1\>_E$ and the joint state $|2\>_S|0\>_E$ evolves to $|2\>_S|2\>_E$. We assume that each of these evolutions is a simple rotation in the joint state space $\H_E\ox\H_S$, lasting for a time $\tau$:
\[
|k\>_S|0\>_E \; \longrightarrow \; |\Psi_k(t)\> = \cos\lambda t|k\>_S|0\>_E + \sin\lambda t|k\>_S|k\>_E
\]
$(k = 1,2; \; 0\le t\le \tau)$ where $\lambda = \pi/2\tau$. At times outside the interval $[0,\tau]$, the joint state of the system and the experimenter is assumed to be stationary (with zero energy). This time development is produced by the Hamiltonian
\[
H = i\hbar\lambda\bigg(|1\>\<1|_S\ox\big[|1\>\<0| - |0\>\<1|\big]_E + 
|2\>\<2|_S\ox\big[|2\>\<0| - |0\>\<2|\big]_E\bigg),
\]
which is switched on at $t = 0$ and off at $t = \tau$.

Suppose the system has just one beable $M$, the observation of the experimenter, with values $(0,1,2)$, and suppose the initial state of the joint system is $\big(a|1\>_S + b|2\>_S\big)|0\>_E$. This has the definite value 0 for the beable $M$, so it is both the real state vector for the joint system and the pilot vector at $t = 0$. Then in the time interval $[0,\tau]$ during which the measurement is proceeding, the pilot state is 
\begin{align*}
|\Psi(t)\> &= a|\Psi_1(t)\> + b|\Psi_2(t)\>\\
&= \cos\lambda t\big(a|1\>+ b|2\>\big)_S|0\>_E + \sin\lambda t\big(a|1\>_S|1\>_E + b|2\>_S|1\>_E\big)
\end{align*}
and the real state of the joint system at any time in this interval is one of the three states $|\Psi(0)\> = (a|1\>_S + b|2\>_S)|0\>_E$, $|1\>_S|1\>_E$ or $|2\>_S|2\>_E$. It can make a transition from $|\Psi(0)\>$ to $|1\>_S|1\>_E$ or to $|2\>_S|2\>_E$ because the (real) matrix elements $(i\hbar)^{-1}\big(\<k|_S\<k|_E\big) H \big(|k\>_S|0\>_E\big)$ ($k = 1,2$) are both positive. It cannot make the reverse transitions because the  
matrix elements $(i\hbar)^{-1}\big(\<k|_S\<0|_E\big) H \big(|k\>_S|k\>_E\big)$ are negative, and it cannot make transitions between $|1\>_S|1\>_E$ and $|2\>_S|2\>_E$ because the relevant matrix elements of $H$ are zero. Thus at time $t = 0$ the real state vector and the pilot vector coincide; between $t = 0$ and $t = \tau$ the pilot vector $|\Psi(t)\>$ changes smoothly but the real state vector remains at its initial value $|k\>_S|0\>_E$ until some undetermined intermediate time at which it changes discontinuously to either $|1\>_S|1\>_E$ or $|2\>_S|2\>_E$ and remains at that value until $t = \tau$. A calculation of the final probabilities from the transition probabilities as given by Bell yields the expected values $|a|^2$ and $|b|^2$.  


To examine the implication \eqref{r implies w}, we will apply this theory to the measurements in the extended Wigner's friend experiment. We will assume that each of the measurements has duration $\tau < 1$ before the time assigned to it (e.g.\ A's measurement ``at time $t = 3$" occupies the interval $[3 - \tau, 3])$, and that each measurement consists of a simple rotation as described above.

 If the result of $F_1$'s measurement at $t = 0$ is $r = $``tail", then the component of $|\Psi(0)\>$ describing the actual world must be $|\Phi(0)\>$. 
The pilot vector is still $|\Psi(0)\>$. $F_1$'s preparation of the electron spin at $t=1$ is accomplished by a unitary operator acting only on $F_1$ and $S$, such that there are no matrix elements of the Hamiltonian between states with different values of the beables $r,x,z,w$; therefore the real state at $t=1$ is $|\Phi(1)\>$. The next measurement, by $F_2$ at $t = 2$, is driven by the Hamiltonian $\1_{F_1C}\ox (H_2)_{F_2S}\ox\1_A\ox \1_W$ where
\begin{multline}
H_2 = i\hbar\lambda\Big(|-\>_{F_2S}\big(\<\downarrow|_S\<0|_{F_2}\big) - \big(|\downarrow\>_S|0\>_{F_2}\big)\<-|_{F_2S}\\
 + |+\>_{F_2S}\big(\<\uparrow|_S\<0|_{F_2}\big) - \big(|\uparrow\>_S|0\>_{F_2}\big)\<+|_{F_2S}\Big).
\end{multline}
The pilot state during the measurement is $\cos\lambda t|\Psi(1)\> + \sin\lambda t|\Psi(2)\>$; the real state must therefore be one of the viable components of $|\Psi(1)\>$ or $|\Psi(2)\>$. Since this Hamiltonian has no matrix elements betweeen states containing $|\head\>_{F_1C}$ and states containing $|\tail\>_{F_1C}$, the only possible transitions from $|\Phi(1)\>$ are to the second or third term in $|\Psi(2)\>$, followed by transitions back to $|\Phi(1)\>$ or to other components of $|\Psi(2)\>$. But the Hamiltonian also has no matrix elements between different viable components of $|\Psi(2)\>$, and the only positive matrix elements of $H/i\hbar$ are those corresponding to transitions in the forward direction, so once a transition has been made to one of the three terms in $|\Psi(2)\>$, there will be no further transitions during this measurement. Thus if the real state after $F_1$'s measurement has $r = $ ``tail", this will still be the case after $F_2$'s measurement and the real state will be the second or third term of $|\Psi(2)\>$, and both of these are possible. Thus there is a non-zero probability that the real state evolves as $|\Psi(0)\> \rightarrow |\Phi(1)\> \rightarrow |\Phi(2)\>$. 

$A$'s measurement of $F_1$ and $C$ at $t = 3$ is driven by the Hamiltonian
$H_3\ox\1_{F_2S}\ox\1_W$ where $H_3$, acting in $\H_{F_1C}\ox\H_A$, rotates $|\fail\>_{F_1C}|0\>_A$ to $|\fail\>_{F_1C}|\fail\>_A$ and $|\ok\>_{F_1C}|0\>_A$ to $|\ok\>_{F_1C}|\ok\>_A$. In terms of the viable states, this is 
\begin{align*}
H_3 &= \half i\hbar\lambda\big(|\head\> + |\tail\>\big)\big(\<\head| + \<\tail|\big)_{F_1C}\ox\big(|\fail\>\<0| - |0\>\<\fail|\big)_A\\
&+ \half i\hbar\lambda\big(|\head\> - |\tail\>\big)\big(\<\head| - \<\tail|\big)_{F_1C}\ox\big(|\ok\>\<0| - |0\>\<\ok|\big)_A.
\end{align*}
This Hamiltonian $H$ has 
\[
\<\Phi(3)|\frac{H}{i\hbar}|\Phi(2)\> > 0,
\]
and there are no positive matrix elements $\<\phi|\tfrac{H}{i\hbar}|\Phi(3)\>$ for viable states $|\phi\>$, so the transition $|\Phi(2)\> \rightarrow |\Phi(3)\>$ is possible, and if it occurs the system remains in the state $|\Phi(3)\>$ until the next measurement.

$W$'s measurement of $F_2$ and $S$ at $t = 4$ is driven by the Hamiltonian $\1_{F_1C}\ox\1_A\ox H_4$ where $H_4$ is the following operator on $\H_{F_2S}\ox\H_W$:
\begin{align*}
H_4 = i&\hbar\lambda|\ok\>\<\ok|_{F_2S}\big(|\ok\>\<0|-|0\>\<\ok|\big)_W\\
+i&\hbar\lambda|\fail\>\<\fail|_{F_2S}\big(\fail\>\<0| - |0\>\<\fail|\big)_W.
\end{align*}
This has
\[
\<\Phi(3)|\frac{H}{i\hbar}|\Phi(4)\> > 0,
\]
and there are no positive matrix elements $\<\phi|\tfrac{H}{i\hbar}|\Phi(4)\>$ for viable states $|\phi\>$, so the transition $|\Phi(3)\> \rightarrow |\Phi(4)\>$ is possible during $W$'s measurement, and if it occurs the system remains in the state $|\Phi(4)\>$. 

Thus it is possible that $W$ and $A$ both get the result ``ok" for their measurements, and this happens even though $F_1$ records the result $r =$ ``tail". This contradicts the theorem of Frauchiger and Renner.

\section{Discussion}

The purpose of this paper has been to show that there is a counter-example to the theorem that Frauchiger and Renner claim to prove. There is a theory which is self-consistent, in which any experiment has only one result, and which reproduces the predictions of quantum mechanics. It is not the purpose of the paper to advocate this theory as a true description of the experiment, but simply to show that it exists. This disproves the theorem. But what is wrong with Frauchiger and Renner's proof?

Let us examine the implication \eqref{r implies w}: if the result of $F_1$'s measurement at $t = 1$ is $r = $ ``tail", then $F_1$ acts on this information and calculates the future development of the whole system by means of the Schr\"odinger equation, with the measurement result ``tail" as initial condition. This is to follow the instructions of the quantum mechanics textbooks, so Frauchiger and Renner describe it as ``compliance with quantum theory". It incorporates a collapse of the state vector on measurement, otherwise known as the collapse postulate. In Bell-Bohmian theory, on the other hand, although the result of measurement determines the real state, the Schr\"odinger equation is applied with a different initial condition, namely the pilot vector. This includes a term corresponding to the result of measurement which did not actually occur. 

Naturally, these two procedures give different results. They are both presented as ``compliant with quantum theory", but this cannot be true if ``quantum theory" has a well-defined meaning. This does not seem to be so. The contradiction between the Frauchiger-Renner claim that ``It is impossible for any theory to obey ({\bf QT}), ({\bf SW}) and ({\bf SW})" and the claim of this paper that ``Bell-Bohmian theory obeys ({\bf QT}), ({\bf SW}) and ({\bf SW})" is due to different meanings of ({\bf QT}) in the two claims.

The version of quantum theory assumed by \FR \ seems appropriate for use by a particular observer, existing as part of the system being described. If $F_1$ at $t = 1$ sees the result ``tail", then it is reasonable for $F_1$ to use the state vector $|\Phi(1)\>$, incorporating this result, to describe the world they are part of. But does this mean that they should use this to calculate what will happen at later times? 

In the Frauchiger-Renner scenario $F_1$ knows that the state vector at $t = 0$ is $|\Psi(0)\>$, which at $t = 1$ has evolved to the state containing a term corresponding to the result of measurement which did not actually occur. $F_1$ is therefore in a position to include this term when calculating what can happen at $t = 5$. 

The rules of ``standard quantum theory", as understood by \FR, are appropriate for use in the more usual situation where the only available knowledge is the result of the experiment. In this situation the only option is to apply the projection postulate. In principle, as the FR experiment shows, the result of such a calculation will be different from one in which the projection postulate is not applied. However, in a realistic experiment with macroscopic apparatus, the difference between the results of the two calculations will be utterly negligible. 

In Bell-Bohmian theory, and in other interpretations of quantum theory, the projection postulate is an approximation which is valid in many circumstances when a quantum system is entangled with a macroscopic system. It is not a fundamental postulate of the theory (it is too ill-defined to be anything of the sort), and there will be situations in which it does not apply. The extended Wigner's friend experiment, as presented by \FR, is one such situation. The dimensions of the system, consisting of a small number of qubits and qutrits, might be small enough to make it possible to realise this experiment. It would be very surprising if the result accorded with a calculation using the projection postulate. 

Each of the agents in the experiment has a different perspective. This will lead them to apply what \FR \ call ``standard quantum theory" in different ways. Calculating at $t = 0$, they will obtain different predictions for the results at $t = 4$. Each of $F_1$, $F_2$ and $A$ will allow for the two possible outcomes of their own measurement, with known probabilities, and calculate the evolution after their measurement as if one result or the other had definitely occurred; that is, they apply the projection postulate to their own measurement while treating the other measurements as purely quantum processes, with no projection. The purely quantum evolution of all the measurements can be regarded as a ``God's eye view" of the experiment. Wigner (who of course is God) makes this calculation, as there is no evolution to be considered after his measurement.

The results of these calculations are as follows. The probabilities of the four possible results of measuring $(x,w)$ at $t = 4$, as calculated by the four agents at $t = 0$, are given in the following table:
\[\label{table}
\begin{array}{l|c|c|c|c|}{}& (\ok,\ok) & (\ok,\fail) & (\fail,\ok) & (\fail,\fail) \\&&&&\\\hline&&&&\\
F_1\quad  &\frac{1}{12}&\frac{5}{12}&\frac{1}{12}&\frac{5}{12}\\&&&&\\\hline &&&&\\
F_2\quad &\frac{1}{12}&\frac{1}{12}&\frac{5}{12}&\frac{5}{12}\\&&&&\\ \hline &&&&\\
A\quad &\frac{1}{4}&\frac{1}{4}&\frac{1}{20}&\frac{9}{20}\\&&&&\\\hline &&&&\\
W\quad &\frac{1}{12}&\frac{1}{12}&\frac{1}{12}&\frac{3}{4}\\&&&&\\\hline
\end{array}
\]

These calculations make no appeal to a ``single-world" assumption. It is only assumed that an observer who sees a result of an experiment sees just one result. This is true, for example, in the ``many worlds" interpretation, in which each world contains just one result of the experiment. The contradiction between the predictions in \eqref{table} comes from the different applications of the rules of standard quantum theory. This appears to show that of the three assumptions {\bf QT}, {\bf SW} and {\bf SC} of \FR, {\bf SW} is not needed to obtain a contradiction: given the meaning they assign to ``standard quantum theory", {\bf QT} by itself is self-contradictory. A similar conclusion has been reached by \cite{GertrudeStein}

The extended Wigner's friend experiment devised by Frauchiger and Renner remains of great conceptual value. It demonstrates that in a single-world theory like Bell-Bohmian theory, possible experimental results which were not realised in the actual world can still have an influence on the future of the actual world. The same moral holds in interpretations of quantum theory which do not postulate a single world in this sense, for example versions of Everett's relative-state theory in which the experience of a sentient physical system is recognised as having its own reality \cite{verdammte}. Events which, for such an observer, might have happened, but didn't, can still affect real future events.

\section*{Acknowledgement}

I am grateful to Renato Renner and Roger Colbeck for illuminating discussions.


\end{document}